\documentclass[11pt,twoside]{article}
\usepackage{asp2010}
\usepackage{graphicx}

\resetcounters

\begin{document}

\title{Gas accretion in disk galaxies}
\author{Francoise Combes}
\affil{Observatoire de Paris, LERMA, CNRS, 61 Av de l'Observatoire, 75014 Paris, France}

\begin{abstract}
Gas accretion is necessary to maintain star formation,
spiral and bar structure, and secular evolution in galaxies.
 This can occur through tidal interaction, or mass
accretion from cosmic filaments. Different processes
will be reviewed to drive gas towards galaxy centers and 
trigger starbursts and AGN.
 The efficiency of these dynamical processes can
be estimated through simulations and checked by
observations at different redshift, across the Hubble time.
 Large progress has been made on galaxies at moderate and high redshifts,
 allowing to interpret the star formation history and star formation
efficiency as a function of gas content, dynamical state and galaxy evolution.
\end{abstract}

\section{Introduction}
 In the last decade, cosmological simulations have emphasized the importance of cold gas accretion
onto galaxies in the mass assembly, in particular at high redshift (e.g. Keres et al. 2005, Dekel et al. 2009,
Devriendt et al. 2010). This mode of accretion is thought to be about one order
of magnitude more important than galaxy mergers in mass assembly, unlike what was assumed in the
hierarchical scenario.

In parallel, simulations of isolated galaxies show how important is the gas accretion to maintain 
star formation at a constant level, as is observed in spiral galaxies, to explain abundance gradients,
and also to maintain the spiral structure. In the following, the impact of gas accretion is first described,
and then we will review the evidence of circumgalactic gas inflow.

\section{Role of gas accretion: secular evolution, bars}
 Secular evolution involves mainly the disk galaxies of the Hubble sequence: all spirals and irregulars.
As for the ellipticals, their formation scenario is still heavily relying on mergers, either a few major mergers,
or more likely a series of minor mergers, to heat the stellar component, destroy disks progressively,
and cancel out any angular momentum, explaining its low values in this class.

In disk galaxies, the main motor of evolution is non-axisymmetries and bars. When the disk is
abundant in gas, as are most high redshift disky objects, then the bars are not long-lived, but
weakened and destroyed by the accumulation of mass in the centers, and by the exchange of angular
momentum between the gas and the stars of the bar (Friedli \& Benz 1993, Berentzen et al. 1998,
Bournaud \& Combes 2002).  A weakened bar would transiently look like a lens inside its
inner ring (e.g. Laurikainen et al. 2009). 

The frequency of bars in disk galaxies as a function of mass shows an interesting
bimodality (Nair \& Abraham 2010), with two maxima in the blue cloud and in the red sequence.
This was also found by Masters et al (2011) with the Galaxy Zoo, although not
by the S4G consortium (see K. Sheth, this meeting), but this could be due to selection effects.

\begin{figure*}[ht]
\centerline{
\includegraphics[angle=-0,width=12cm]{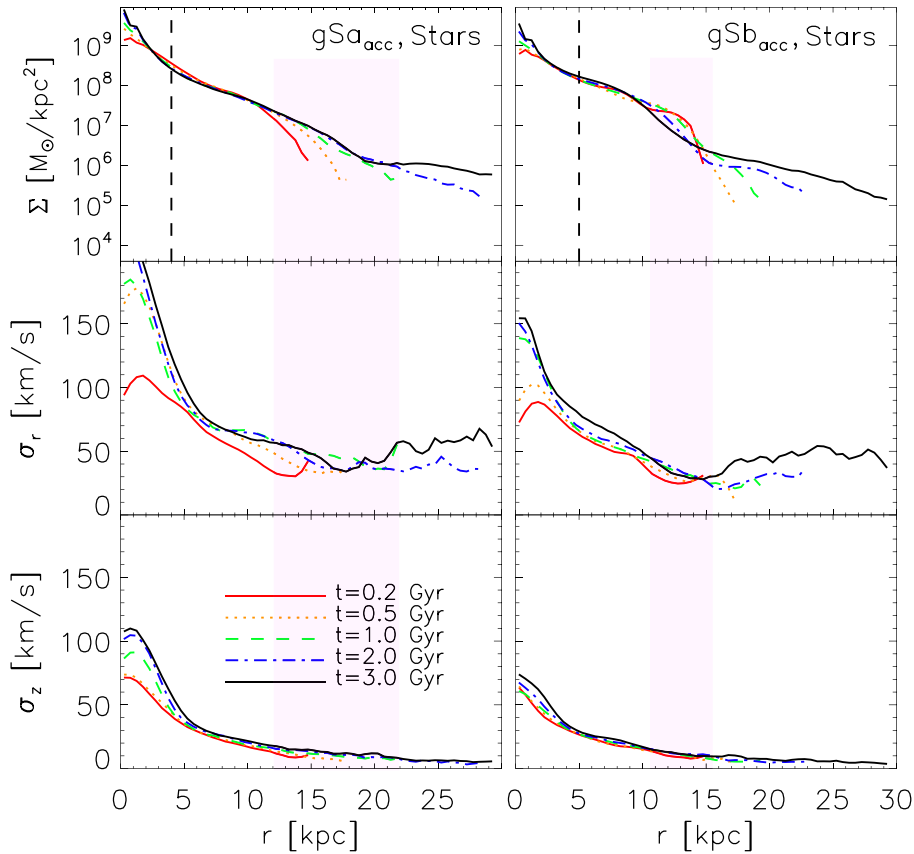}}
\caption{Simulations of smooth, in-plane gas accretion, for two spiral galaxies,
giant Sa (left) and giant Sb (right).  The various colored lines correspond
to different epochs. The first row shows the evolution of stellar surface
density, and the two others the radial and vertical velocity dispersions.
From Minchev et al (2012).}
\label{fig:minchev-12}
\end{figure*}

Bars can act in conjunction with spirals, to redistribute the angular-momentum across
galaxies, and to modify significantly stellar radial profiles. The non-linear interactions 
at resonances overlap multiply the effects (Minchev et al 2011).
In less then 3 Gyrs, the effective sizes of galaxy disks may be multiplied by 3, and the
corresponding radial migration brings high-velocity dispersion stars in the outer parts.
  Disk thickening is also substantial (Minchev et al 2012). 
When gas accretion is considered, the strength of bars can be re-boosted, and new stars
formed out of the accreted gas continuously re-shape the radial stellar profiles.
It is then possible to obtain the three observed types: Type I as a single exponential disk, Type II as
a truncated one, when star formation is newly forming at the break, and beyond the break the 
star formation threshold is not yet reached, and Type III, the anti-truncated profile can be 
obtained in case of strong gas accretion in the outer parts (cf Fig \ref{fig:minchev-12}).  The Type III morphology appears
relatively transient, and able to evolve into Type II or Type I.  Its presence could be a tracer of accretion
events. There is some correlation of these Type II with weakened bars, as expected from
strong gas accretion (Bournaud \& Combes 2002, Combes 2011).

When the gas accretion occurs essentially from non-aligned cosmic filaments,
characteristic signatures may occur, such as inclined and warped rings (Roskar et al 2010),
or even polar rings, when the accretion is near polar. Brook et al (2008) have shown how
a lenticular system is first formed from matter accretion, which suddenly stops when 
the birth filament is consumed out.  The next filament in the perpendicular direction
then fuels a relatively stable polar ring system.
 Gas accretion may mimick galaxy interactions, since it can produce asymmetries,
lopsidedness, clumpiness, and sfatbursts. Even if the accretion is globally symmetric and isotropic, 
it may be temporarily on one side only.

Gas accretion replenishes the extended gas reservoirs present around most spiral galaxies. This gas slowly
 spirals in, when in quiescent state. However, the first tidal interaction may drive the gas violently towards
the center, and strongly affects abundance gradients, that can even be reversed
(Montuori et al 2010): low-metallicity gas flows into the center and dilutes the abundance
of the central gas, in a time-scale shorter than the time required for this gas to re-enrich through the 
triggered nuclear starburst.  Such gradient reversals have been observed at high redshift
in the MASSIV survey (Queyrel et al 2012).

\section{Inside-out disk formation, inflow/outflow, metallicity}

Gas accretion occurs in the outer parts of disks, and is a way to explain inside-out disk formation.
There are now multiple evidence of this progressive mode of disk formation.
 Through observations of galaxies as a function of redshift, it is possible to track
the evolution directly. However, finding the progenitors of today galaxies is not easy.
Statistically, this is solved by matching the galaxies at a given cumulative number density,
for instance 1.4 10$^{-4}$ Mpc$^{-3}$. Plotting the mass of these matched galaxies at a given 
redshift, Patel et al (2013) follow the mass increase of galaxies over z=3 to 0. They notice that
this mass increase involves only the mass of the outer parts, while the mass inside 2kpc is stable in all
the same galaxies. 
This fact is related to the observations that the normalised galaxy radius,
at a given mass, increases by a factor 3 from z=3 to 0 (Newman et al 2012).
It is possible that dry minor mergers explain this increase of size,
without much mass accretion, in the case of quenched galaxies. For star forming 
disk galaxies, external gas accretion, accompanied by secular evolution, is necessary.
In the GASS sample of local galaxies detected in HI-21cm, Wang et al (2011) find that galaxies
richer in HI are bluer, and the more so in the outer parts: the radial gradient of color
is a function of HI-mass fraction. They conclude that the gas must be accreted slowly, and relaxed,
since there is no correlation between HI-fraction and lopsidedness in this sample.

More directly, looking at H$\alpha$ in the 3D-HST project in 57 galaxies at z$\sim$ 1, Nelson et al (2012)
find that the effective radii of galaxies in ionised gas and new stars is about 1.3 larger
than the effective radius of the rest-frame R-band stellar continuum, representing older stars. The effet
is larger for massive galaxies, that certainly are the first to form inside-out.

In this inside-out scenario,  with gas accretion, the radial migration will produce a typical
reversal of stellar ages in the outer parts: the gas is accreted at the radius of the break,
which is where the new stars are formed, accentuating the negative age gradient from the center.
After the break, only old stars migrated from the center are expected, and there is now a positive
gradient (Roskar et al 2008).  This age gradient reversal has been observed in M33 by
Williams et al (2009). Some other galaxies do not show any break, but a flat age 
gradient in the outer parts (Vlajic et al 2011).

\section{Evidence of gas accretion: HVC, warps, QSO absorption}

Most of the gas from cosmic filaments is accreted at large scales, settles down to the disk,
and spirals in progressively. Since the alignment process occurs through precession and dissipation,
with a time-scale of the order of a few dynamical times at these large radii, this can take some
Gyrs, during which galaxies appear warped or perturbed in the outer parts.
Warps and polar rings are therefore the best tracers of external accretion, and indeed
most spiral galaxies are observed to be warped (e.g. Briggs 1990, Binney 1992, Reshetnikov \& Combes 1998).
Also the frequency of asymmetries and lopsidedness in spiral galaxies cannot
be explained but with external accretion (Jog \& Combes 2009).

Searches have been done of extra-planar gas in the halo of spiral galaxies
(Fraternali et al. 2002, Heald et al 2011, Gentile et al. 2013)
 and the quantities found are relatively small, NGC 891 being the most remarkable for
its gas entension. The origin of this gas is multiple. 
Some gas can be ejected into the halo by stellar feedback (fountain effect), or through
tidal disruption of satellites. In the Milky Way, the High Velocity Clouds (HVC)
and the Magellanic stream are good examples. This 
gas is accreted progressively, with an interface
 of multiphase gas, but at a rate lower than the star formation rate
(0.4 M$\odot$/yr in the Galaxy, Putman et al. 2012).

\begin{figure*}[ht]
\centerline{
\includegraphics[angle=-0,width=12cm]{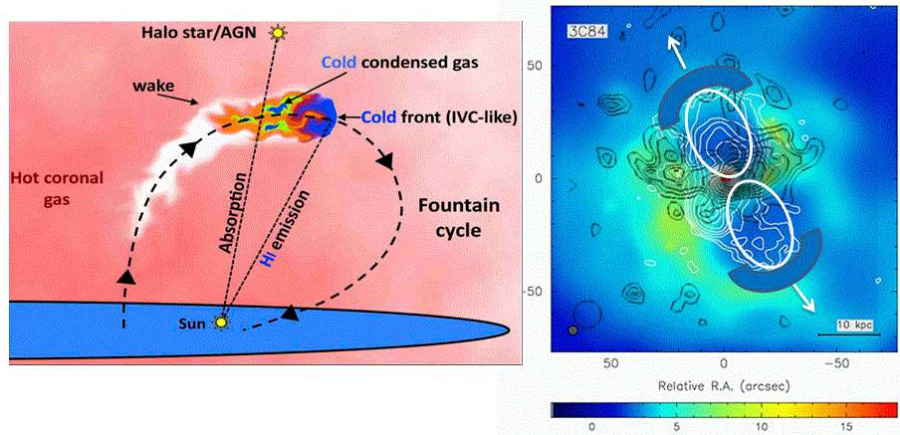}}
\caption{ {\it Left:} Schematical view of the fountain gas in the hot corona of the Milky Way, from 
Fraternali et al (2013).  Ionised gas is elevated to the hot halo, compresses and provokes the cooling of 
the hot gas, which comes back to the plane, as high (or intermediate) velocity clouds. This is a way to accrete external gas, 
since more gas cools down than was elevated by the star formation feedback.
{\it Right:} An analogous feedback mechanism occurs in cool core clusters. Here we see in the Perseus cluster,
the two radio jets filling two cavities in the hot X-ray gas (white contours and bubbles), where the heated X-ray gas
does not cool. At the boundary of the cavities, gas can cool (enhancement of X-ray emission), and even very cold gas
forms and is visible in CO emission (dark contours and arcs). In this scenario, gas is not cooling towards the center,
as previously expected, but at $\sim$ 20kpc form the center of the cluster.  
}
\label{fig:CF-halo}
\end{figure*}

The way external hot gas is accreted might be complex (Fraternali \& Binney 2008).
The fountain effect ejects ionised gas into the halo, where it encounters the hot coronal gas.
Merging with this assumed non-rotating gas, it looses angular momentum, and cold gas condensates
in the shock. Finally more gas is infalling down, that was uplifted by star formation feedback.
The metallicity of the infalling gas is relatively high, since it is a mixture of gas from very different
origins.
 The same kind of processes is invoked in cool core clusters, as schematically shown in Figure  
\ref{fig:CF-halo}. The hot X-ray gas in the cluster halo is dense enough at the center to cool down
and fuel a central AGN. But the radio jets of the AGN re-heat the medium, in creating two
cavities of plasma, pushing the X-ray gas out at the cavity boundaries. There the gas is cooling down
to low temperatures, and molecular clouds are observed through their CO emission (Salome et al 2006, 2008).
 In parallel, the AGN jets drag some molecular gas previously settled in the central galaxy, and this uplifts some
high-metallicity gas at 20-30kpc, which explains the more efficient cooling in filaments far from the center,
and the gas abundance sufficient to produce CO emission.

High velocity clouds have been found around M31 and M33 in the local group, and also 
along a gas bridge between M31 and M33 (Lockman et al. 2012).  About 121 HVC have
been detected around the isolated galaxy NGC 6946 (Boomsma et al. 2008). In nearly face-on galaxies,
these HVC re traced by the numerous HI holes they created in the disk plane, as demonstrated in 
M101 (Kamphuis et al. 1991). Some of these highly perturbed galaxies are aslo lopsided,
like UGC7989 (Noordermeer et al. 2005). More generally, 30\% of galaxies show
an asymmetry larger than 10\%, as quantified by the Fourier decomposition of the density.
In poor environments, low surface brightness (LSB) galaxies are dwarfs particularly rich in HI gas,
and revealing spectacular warps, like NGC 2915 (Meurer et al. 1006), 
NGC 5055 (Battaglia et al. 2006), or NGC 3741 (Begum et al. 2005, Gentile et al. 2007).

Another important way to discover the circum-galactic gas around galaxies is through absorption 
measurements in front of remote quasars. The various systems can be sorted by their
column density, from the Ly$\alpha$ forest at very low columns, to the damped Ly$\alpha$ (DLA)
at NHI $>$ 10$^{20}$ cm$^{-2}$, passing through the Lyman limit systems (LLS). The abundance
of the systems varies as a power law (Prochaska et al. 2010). Fumagalli et al. (2011) have derived
from cosmological simulations the probability to observe absorptions in front of backgound sources, at z=2.3, and 
according to the distance from the center of the galaxy up to the virial radius. The result is that there are
only 10\% of line of sights optically thick in HI, and 1\% probability to find a DLA. About 30\% of
absorbers come from massive galaxies and their streams. There has been some debate about the filling
factor of gaseous filaments around galaxies. While Dekel et al. (2009) claim a filling factor
for cold filaments of 25\% between 20 and 100kpc around a halo of total mass 10$^{12}$ M$_\odot$ at z=2.5,
in the same conditions, Faucher-Gigu\`ere \& Keres (2011) or Kimm et al. (2011) find 2\% or 5 \% 
respectively. So the observation of cold gas accretion is difficult, for the small
filling fators, but also due to the confusion with the galaxy host, when the line-of-sights are too close,
or the low metallicity of the circum-galactic gas, if carbon lines are used.

Another possibility would be to detect Ly$\alpha$ photons emitted when the gas is infalling in 
dark matter haloes, in some way when it re-radiates its gravitational energy.
It has been proposed that the frequently observed Ly$\alpha$ blobs at high redshift are precisely
due to this radiation. But the simulations from Faucher-Gigu\`ere et al. (2010),
taking into account the proper self-shileding, etc.,  have shown that the gravitational emission
in these blobs is negligible.  Ly$\alpha$ blobs are powered by star formation or AGN.
Now, several hundreds of Ly$\alpha$ blobs have been discoverd (Matsuda et al. 2006, 2011).
 The Ly$\alpha$ line is resonant and requires radiative transfer to better understand the shapes
of the profiles observed. Verhamme et al. (2006) have showned that in case of outflowing material,
a P-cygni profile is expected, with emission on the red side, and absorption on the blue side.
The reverse is expected for inflowing gas. However, in all Ly$\alpha$ blobs observed until now, there
are always P-cygni profiles with emission in the red, therefore only outflows have
been detected.

Another way to detect the filaments could then be through fluorescence of Ly$\alpha$ photons emitted by 
a starburst or a luminous quasar. This has been done in a few cases
(Rauch et al. 2011, Cantalupo et al. 2012), but this clever technique should be developed more. A powerful
quasar can illuminate dark gas, or dark galaxies, up to 100kpc distance.

\begin{figure*}[ht]
\centerline{
\includegraphics[angle=-0,width=12cm]{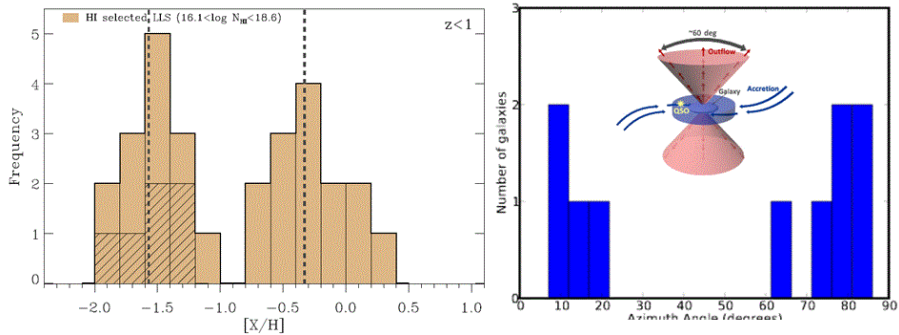}}
\caption{ {\it Left:} Bimodal metallicity distribution of the Lyman Limit Systems (LLS) at z$<$1. The hashed histograms 
correspond to upper limits. The peak of the tw maxima are marked by vertical dotted lines. The low metallicity LLS could
correspond to inflows, while the high metallicity ones to outflows. From Lehner et al (2013).
{\it Right:} Bimodal distribution of the MgII absorbers at z$\sim$0.1, according to the angle from the major axis of the hosts.
 According to the small drawing, it is expected that outflows occur on the minor-axis, and inflow on the major-axis.
From Bouch\'e et al. (2012). 
}
\label{fig:Flows-stat}
\end{figure*}

 Finally, when absorption lines are detected in front of quasars, what are the methods
to distringuish inflows from outflows in the circumgalactic medium (CGM)?
Two methods have been used, and are illustrated in Figure \ref{fig:Flows-stat}.

Using an H I-selected sample of 28 Lyman limit systems (LLS) at z$<$1, observed in absorption with the HST-COS
spectrograph, Lehner et al. (2013) are able to determine their metallicity from weakly ionized metal species (e.g., O II, Si II, Mg II) 
and find that  the metallicity distribution of the LLS is bimodal with metal-poor and metal-rich branches peaking 
at about 2.5\% and 50\% solar metallicities.  Both branches have comparable number
of absorbers.  The metal-rich branch likely traces winds, recycled outflows, and tidally stripped gas,
while  the metal-poor branch has properties consistent with cold accretion streams 

When the galaxy host is detected in the proximity of the 
absorbant lines of sight, it is possible to determine the azimuthal orientation of the gas:
Is it towards the minor axis? more likely to be an outflow, or the major axis?
then it is probably an inflow. Bouch\'e et al. (2012, 2013) have shown that the number of absorbants
as the function of angle with respect to the major-axis, reveals a bimodal distribution also. These are MgII absorbants
at z$\sim$ 0.1 (Fig   \ref{fig:Flows-stat}). 
The outflow speeds are found to be 150-300 km/s, i.e. of the order of the circular velocity, and smaller than the escape 
velocity by a factor of $\sim$2. The outflow rates are typically two to three times the instantaneous SFRs.

\section{Evolution with redshift of gas content}

It is now well established that the global SFR in the Universe evolves towards a maximum
at about z=1-2, and then decreases by a factor 10 to the present (e.g. Bouwens et al. 2012).
When the SFR is normalized to the stellar mass, the specific SFR (called sSFR) is derived,
which is the inverse of the growth time-scale for the galaxy. This typical time-scale has been 
establied to be 2 Gyr at z=0, and then regularly decreases with increasing z. There was a recent debate
about a possible plateau of the sSFR after z=2, but this has been corrected (Smit et al 2013).
The first galaxies are so active in forming stars, that their nebular emission dominates their total emission.
Without precise spectroscopy, it was not possible to disentangle the continuum from the lines, and the
continuum was over-estimated. With the recent re-normalisation, the sSFR increases
at all redshift, which corresponds better to the simulations predictions.

Why galaxies in the main sequence of star formation are more efficient to form stars at high redshift?
According to our PHIBSS survey, detecting about 52 galaxies at z=1.2 and 2.3
with the IRAM interferometer, we determine that the gas fraction over all baryons in these galaxies 
increase with redshift up to 50\% at least. The gas fraction is 34\% at z=1.2 and 44\% at z=2.3, while
it is only of the order of 5\% at z=0 (Tacconi et al. 2010, 2013).

\begin{figure*}[ht]
\centerline{
\includegraphics[angle=-0,width=10cm]{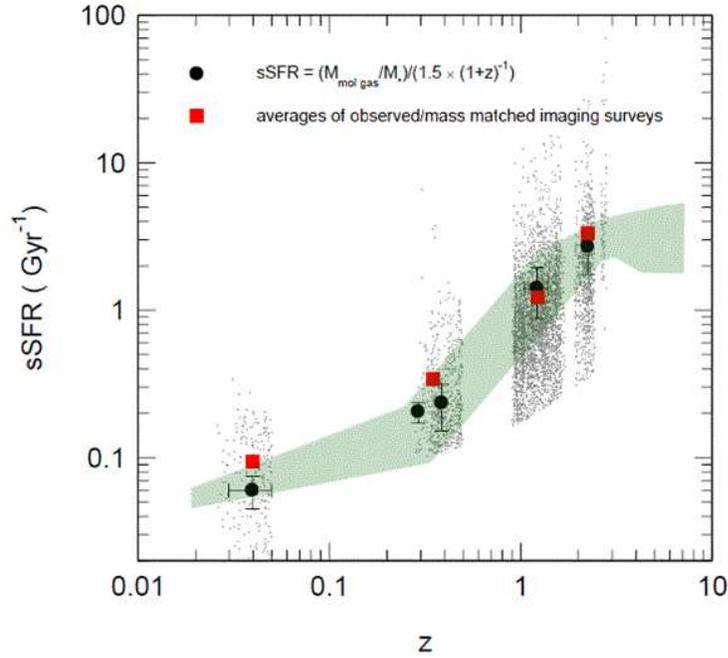}}
\caption{ Specific star formation rate sSFR=SFR/M$_*$  as a function of redshift. 
The large blue filled circles denote the sSFR inferred from the average CO gas fractions, 
using t$_{dep}$=Mmol gas/SFR=1.5  (1+z)-1.05 Gyr inferred from the 
COLDGASS and PHIGS surveys (1.5 Gyr at z=0 and 0.64 Gyr at z=1.2). 
The grey dots are individual galaxies from SDSS and imaging surveys,
mass-matched to the CO-surveys, and the crossed red squares denote the average sSFR derived from 
these data points in the four
redshift slices of the CO surveys. The green shaded region are estimates of mean sSFR as a function of redshift from 
various optical/infrared  imaging surveys in the literature, as compiled in Weinmann et al. (2011), 
Sargent et al. (2010) and Gonzalez et al (2012). 
}
\label{fig:phibss}
\end{figure*}

The evolution of the specific SFR is shown in Figure \ref{fig:phibss}, where we have assumed that the depletion time-scale for star 
formation in the main sequence can be modelled as t$_{dep}$ = 1.5 (1+z)-1.05  Gyr. We have shown that the gas fraction
was strongly correlated to the sSFR, and explored the whole range from 10 to 90\%. In the Kennicutt-Schmidt diagram,
the surface density of gas and the SFR surface density follows the main sequence branch. With respect to the gas surface density
divided by the dynamical time, all objects (main sequence and starbursts) align on the same almost linear curve.
This is due to the smaller dynamical time-scale of nuclear starbursts.  

When only ULIRGS and starbursts are considered, the strong increase of star formation efficiency rate
is explained at high redshift by two factors with comparable weight:
first the gas fraction is higher by a factor 3 at z=1 with respect to z=0, and the star formation efficiency  (SFR per unit gas mass),
is also higher by a factor 3 (Combes et al 2013).

\section{Conclusion}

External gas supply is fundamental for secular evolution: bars drive gas towards the very center,
and accretion is necessary to replenish the disk, to reform a bar, or for the galaxy to stay
on the blue cloud, instead of directly move on the red sequence.

The gas accretion is slow, progressive, from gas reservoirs in the outer parts
of galaxies. Matter and angular momentum is redistributed by
non-axisymmetries and bars, including radial migration.
These processes explain the inside-out disk formation, that is now
currently observed at all redshifts.
The evolution of the disk sizes at a given mass can also be explained
by slow matter accretion and dry minor mergers. In particular secular evolution can 
multiply the effective radii by up to a factor 3.

Gas accretion was more intense in the past. It could be the explanation,
together with galaxy interactions and fly-bys, to abundance gradients reversal
in galaxy disks. Tidal forces drive the gas reservoir into the center, and this nearly
primordial gas dilutes the metallicity of the central gas, on a time-scale shorter, than the
enrichment time-scale due to the triggered starburst.

Warps and polar rings might be the best evidence of gas accretion,
since their setlling takes a few Gyr, and they reflect non-aligned accretion
of matter.
Other evidence of accretion can be seen in high velocity clouds, high-column density absorbants
in front of quasars. There is a bimodality between the gas inflowing and outflowing,
with almost equal weights, through the metallicity and the angle of accretion.

There is a strong redshift evolution of gas fraction in galaxies, which can explain
the SFR history, peaking at z=1-2. The disk dynamics at high z is different from
what is observed locally, since the high gas fraction makes disks highly unstable and
turbulent.

\acknowledgements Thanks to Mark Seigar and the organisers for such an interesting and nicely
located meeting.   


\end{document}